\documentclass[prc,twocolumn,superscriptaddress,showpacs]{revtex4}
\usepackage[dvips]{graphics} 
\usepackage{graphicx}
\usepackage{longtable}
\usepackage{subfigure}
\usepackage{epsfig}
\usepackage{natbib}
\usepackage{color}
\usepackage[colorlinks = true,linkcolor = blue,citecolor = blue,urlcolor = blue,filecolor = blue,pagecolor = blue, breaklinks=true]{hyperref}
\usepackage{amssymb}
\usepackage{amsmath}

\begin{document}

\title{Novel technique to extract experimental symmetry free energy information of nuclear matter }

\author{J.~Mabiala}
\altaffiliation{Present address: INFN, Laboratori Nazionali di Legnaro, Italy; justin.mabiala@lnl.infn.it}
\affiliation{Cyclotron Institute, Texas A$\&$M University, College Station, Texas 77843, USA}
\author{H.~Zheng}
\altaffiliation{Present address: Laboratori Nazionali del Sud, INFN, via Santa Sofia, 62, 95123 Catania, Italy}
\affiliation{Cyclotron Institute, Texas A$\&$M University, College Station, Texas 77843, USA}
\affiliation{Physics Department, Texas A$\&$M University, College Station, Texas 77843, USA}
\author{A.~Bonasera}
\affiliation{Cyclotron Institute, Texas A$\&$M University, College Station, Texas 77843, USA}
\affiliation{Laboratori Nazionali del Sud, INFN, via Santa Sofia, 62, 95123 Catania, Italy}
\author{P.~Cammarata}
\altaffiliation{Present address: Analytical Technology Center, Dow Chemical Company, Freeport, TX, 77541}
\affiliation{Cyclotron Institute, Texas A$\&$M University, College Station, Texas 77843, USA}
\affiliation{Chemistry Department, Texas A$\&$M University, College Station, Texas 77843, USA}
\author{K.~Hagel}
\affiliation{Cyclotron Institute, Texas A$\&$M University, College Station, Texas 77843, USA}
\author{L.~Heilborn}
\affiliation{Cyclotron Institute, Texas A$\&$M University, College Station, Texas 77843, USA}
\affiliation{Chemistry Department, Texas A$\&$M University, College Station, Texas 77843, USA}
\author{Z.~Kohley}
\altaffiliation{Present address: National Superconducting Cyclotron Laboratory, Michigan State University, East Lansing, Michigan 48824, USA}
\affiliation{Cyclotron Institute, Texas A$\&$M University, College Station, Texas 77843, USA}
\affiliation{Chemistry Department, Texas A$\&$M University, College Station, Texas 77843, USA}
\author{L.~W.~May}
\affiliation{Cyclotron Institute, Texas A$\&$M University, College Station, Texas 77843, USA}
\affiliation{Chemistry Department, Texas A$\&$M University, College Station, Texas 77843, USA}
\author{A.~B.~McIntosh}
\affiliation{Cyclotron Institute, Texas A$\&$M University, College Station, Texas 77843, USA}
\author{M.~D.~Youngs}
\affiliation{Cyclotron Institute, Texas A$\&$M University, College Station, Texas 77843, USA}
\author{A.~Zarrella}
\affiliation{Cyclotron Institute, Texas A$\&$M University, College Station, Texas 77843, USA}
\affiliation{Chemistry Department, Texas A$\&$M University, College Station, Texas 77843, USA}
\author{S.~J.~Yennello}
\affiliation{Cyclotron Institute, Texas A$\&$M University, College Station, Texas 77843, USA}
\affiliation{Chemistry Department, Texas A$\&$M University, College Station, Texas 77843, USA}

\date{\today }

\begin{abstract}
A new method of accessing information on the symmetry free energy from yields  of fragments produced in Fermi-energy heavy-ion collisions is proposed. Furthermore, by means of quantum fluctuation analysis techniques, correlations between extracted symmetry free-energy coefficients with temperature and density were studied. The obtained results are consistent with those of commonly used isoscaling techniques.
\textcolor{blue}{}
 
\end{abstract}

\maketitle

\section{INTRODUCTION}
The fundamental goal of  studying strongly excited nuclei is to characterize the equation of state (EOS) of nuclear matter over a wide range of temperature, density, pressure and isospin. This quest to characterize nuclear phenomena at intermediate energies is still a matter of intensive investigations. The characterization of the EOS of nuclear matter plays a key role for various phenomena in nuclear astrophysics, nuclear structure, and nuclear reactions \cite{Lattimer23042004,li2008recent,Janka200738,Danielewicz22112002,Giuliani2014116}. The EOS for asymmetric nuclear matter is usually expressed as a term related to symmetric matter and a term which takes into account the isospin asymmetry of the system. The latter is referred to as the symmetry energy. In early studies, the symmetry energy coefficient  ($E_{sym}$) of nuclei was extracted by fitting the binding energy in their ground state with various versions of the liquid drop mass formula. The properties of nuclear matter are afterwards determined by theoretically extrapolating the nuclear models designed to study the structure of real nuclei which are cold ($T=0$), nearly symmetric ($N\approx Z$) and at the saturation density ($\rho_0\approx$0.16 fm$^{-3}$). In contrast to the value of $E_{sym}$ at $\rho_0$ and $T=0$, the behavior of  $E_{sym}$  with temperature ($T$) and density ($\rho$) is still being mapped out. 

Many experimental and theoretical investigations have been devoted, in recent years, to estimate the behavior of $E_{sym}$ as a function of $T$ and $\rho$. Among these efforts, measurements of the giant dipole \cite{PhysRevC.77.061304}, pygmy dipole \cite{PhysRevC.76.051603} and giant monopole \cite{PhysRevLett.99.162503} resonances in neutron-rich nuclei, neutron-proton emission ratios \cite{PhysRevLett.97.052701}, isospin diffusion \cite{PhysRevLett.92.062701}, collective flows \cite{PhysRevC.85.064605} and fragment isotopic ratios \cite{PhysRevLett.85.716,PhysRevLett.86.5023,PhysRevC.64.054615,PhysRevC.74.024605} have provided constraints on the density dependence of the symmetry energy at subsaturation densities. Recently, a large body of experimental data from studies of heavy-ion collisions \cite{PhysRevC.75.014601,PhysRevLett.104.202501,PhysRevC.85.064618,PhysRevLett.108.172701} have been used to extract the free symmetry energy and the symmetry energy at subsaturation densities and moderate temperatures. In those studies, isoscaling parameters deduced from isotopic yields measured in two similar reactions with different isotopic composition were used  to access the symmetry free energy coefficients. The symmetry energy coefficients were in turn derived using model calculated symmetry entropies together with experimental symmetry free energy coefficients.  

To a good approximation, the EOS of asymmetric nuclear matter can be written as
\begin{equation}
E(\rho,T,m)=E(\rho,T,m=0)+E_{sym}(\rho,T)m^2+O(m^4)\ ,
\label{Eq1}
\end{equation}
where $\rho=\rho_p+\rho_n$, $m=(N-Z)/A$ is the neutron-proton asymmetry and $E(\rho,T,m=0)$ is the EOS of symmetric nuclear matter \cite{PhysRevC.69.064001,PhysRevC.75.014607}. As a general representation of the symmetry energy coefficient, the following definition has been considered 

 \begin{equation}
E_{sym}(\rho,T)=\dfrac{1}{2}\left[E(\rho,T,1)+E(\rho,T,-1)\right]-E(\rho,T,0)\ .
\label{Eq2}
\end{equation}

We recall here that in most cases, the symmetry energy is connected to the isotopic yields through the relationship
\begin{equation}
\alpha=\dfrac{4C_{sym}}{T}\left[\left(\dfrac{Z_1}{A_1}\right)^2-\left(\dfrac{Z_2}{A_2}\right)^2\right]\ ,
\end{equation}
where $C_{sym}$ is the symmetry energy coefficient, $Z_1$, $A_1$ and $Z_2$, $A_2$ are respectively the charge and the mass numbers of system 1 and 2 (system 2 being richer in neutrons than system 1), and $T$ is the common temperature of the two systems \cite{PhysRevLett.86.5023,PhysRevC.74.024605,PhysRevC.65.044610}. However, at sufficiently low densities, $C_{sym}$ was shown to be substituted for the symmetry free energy ($F_{sym}$) and is related to $E_{sym}$ through $E_{sym}=F_{sym}+TS_{sym}$, with $S_{sym}$ the symmetry entropy \cite{PhysRevC.75.014601,PhysRevLett.104.202501,PhysRevC.85.064618,PhysRevLett.108.172701}. This implies that at low densities the symmetry entropy contribution to $E_{sym}$ becomes significant as clustering increases the binding energy and therefore reduces the entropy in symmetric matter \cite{Horowitz200655}.

\section{FORMALISM}
Heavy-ion collisions in the Fermi-energy domain are dominated by nuclear fragmentation and their studies provide information about the properties of nuclear matter at moderate temperatures and sub-saturation densities. Several studies \cite{D'Agostino1999329,PhysRevC.71.054606,PhysRevC.74.054608} have shown this energy domain to be the region of the nuclear liquid-gas phase transition. In our recent works \cite{PhysRevLett.101.122702,PhysRevC.81.044618,PhysRevC.83.054609,S021830131250019X,J.Phys.Conf.Ser.420.012110,PhysRevC.87.017603}, we have analyzed fragment yield data to investigate the nuclear phase transition using the Landau free-energy approach \cite{K.Huang,Laundau.Lifshitz}. In such an approach, the key assumption is that in the vicinity of the critical point, the fragment free energy per nucleon ($F$) relative to the system temperature ($T$) can be expanded in a power series in the fragment's neutron-proton asymmetry $m$ as given by the relation

\begin{equation}
\dfrac{F}{T}=\dfrac{1}{2}am^2+\dfrac{1}{4}bm^4+\dfrac{1}{6}cm^6-\dfrac{H}{T}m\ ,
\label{Eq3}
\end{equation}
where $m=(N_f-Z_f)/A_f$, and $N_f$, $Z_f$, and $A_f$ are the neutron, proton, and mass numbers of the fragment, respectively. The quantity $m$ behaves as an order parameter, $H$ is its conjugate variable and the coefficients $a$, $b$, and $c$ are fitting parameters. We recall that based on a modified Fisher model \cite{Fisher1967,Minich1982458,PhysRevLett.101.122702}, fragment yields are proportional to $A_f^{-\tau}e^{-(F/T)A_f}$ near the critical point; with $\tau$ as the critical exponent.

From Landau's free energy equation, using an analogous expression of Eq.~\ref{Eq2} for $F_{sym}$ (composed of pure neutron, pure proton, and symmetric nuclear matter), one can obtain the following expression:
\begin{equation}
\dfrac{F_{sym}}{T}= \dfrac{1}{2}a+\dfrac{1}{4}b+\dfrac{1}{6}c\ .
\label{Eq4}
\end{equation}
The parameters of the Landau's equation are related to the state variables of the fragmenting system and have been shown to depend on its proton-neutron concentration and excitation energy \cite{S021830131250019X,PhysRevC.87.017603}. This suggests that these parameters could be used to directly obtain information about the symmetry free energy which is a component of the nuclear EOS.

In this paper, we report on experimental symmetry free energy coefficients extracted using the Landau free-energy approach. The temperature and density of the fragmenting source are determined using the quantum-fluctuation method, fully described in Refs.~\cite{Zheng:2010kg,PhysRevC.86.027602,Zheng201243,PhysRevC.88.024607,JPGNuclPartPhys.41.055109}. This is the first time that experimental fragment yield data analyzed within the Landau free-energy framework are used to determine symmetry free energy coefficients in a completely self-consistent manner.

\section{EXPERIMENTAL DETAILS AND EVENT SELECTION}
The experiment was performed at the K-500 superconducting cyclotron facility at Texas A$\&$M University. Beams of $^{64}$Zn, $^{70}$Zn and $^{64}$Ni at 35 MeV/A were used to respectively irradiate $^{64}$Zn, $^{70}$Zn and $^{64}$Ni targets. The 4$\pi$ NIMROD-ISiS array \cite{Wuenschel2009578,Schmitt1995487} was used to collect charged particles and free neutrons produced in the reactions. More details of the experiment have been given in Refs.~\cite{KohleyPhD,PhysRevC.83.044601, PhysRevC.86.044605}. An excellent energy resolution was achieved, allowing isotopic resolution of charged particles up to Z = 17 and elemental resolution up to the charge of the beam. 

We reconstruct the primary hot nuclear system in order to select events of similar character.  For this analysis, we are interested in studying equilibrated systems, so events are selected in the following way: Fragments that do not originate from an equilibrated quasiprojectile (QP) source were excluded with the condition that the longitudinal velocity of fragments with $Z=1,2,\geq 3$ be in the range of $\pm65\%$, $\pm60\%$, and $\pm40\%$, respectively, of the velocity of the heaviest fragment in the event. The sum $Z$ is selected from 21 to 30, the sum $A$ of fragments is selected to be 54 to 64. To select roughly spherical events, which are spatially equilibrated, the quadrupole moment $Q_{shape}$ was required to be $-0.3\leq\log_{10}(Q_{shape})\leq 0.3$ where $Q_{shape}=\sum p^2_z/\sum \frac{1}{2}p^2_t$, and $p_z$ and $p_t$ are respectively the longitudinal and transverse momenta of the fragments comprising the QP. 

The QP system was reconstructed from events in which all charged particles and free neutrons were isotopically identified. The neutron ball provided event-by-event experimental information on the free neutrons emitted during a reaction. The number of free neutrons emitted by the QP was deduced from the total measured number of neutrons, background, and efficiencies for measuring neutrons produced from QP and quasitarget sources \cite{Wuenschel20101,Marini201380}. In order to minimize contributions from collective effects, which are predominant in the beam direction, the excitation energy of the reconstructed QPs was estimated from transverse kinetic energy of the charged particles, the neutron multiplicity, the average neutron kinetic energy determined using the Coulomb-shifted proton energy distribution, and the energy needed for the breakup ($Q$-value). A detailed description of this method of reconstruction is given in Refs.~\cite{PhysRevC.79.061602,Wuenschel20101}. Events were sorted in 8 excitation energy bins, 1 MeV/A wide, from 2.5 to 9.5 MeV/A

\section{TEMPERATURE AND DENSITY}
The temperatures of reconstructed QP sources are obtained with the quadrupole momentum fluctuation method. The method is described in a very detailed way in \cite{Zheng:2010kg,PhysRevC.86.027602,Zheng201243,PhysRevC.88.024607,JPGNuclPartPhys.41.055109,IntJModPhysE.22.1350090}; here we only briefly outline it. The quadrupole momentum is defined as $Q_{xy}=p^2_x-p^2_y$ using the transverse components $p_x$ and $p_y$ of the particle's momentum in the frame of the QP source. If the correct quantum distribution for fermions is used, the variance of $Q_{xy}$ is related to the temperature by
\begin{equation}
\langle\sigma^2_{xy}\rangle = 4m_{part}^2T^2F_{QC}\ ,
\label{Eq5}
\end{equation}
where $m_{part}$ is the mass of the particle being used as the probe and $F_{QC}$ is the quantum-correction factor. The quadrupole is defined in the transverse direction in order to minimize nonequilibrium effects which may manifest in the beam direction. Equation~\ref{Eq5} was solved numerically for a Fermi gas and the quantity $F_{QC}$ was parametrized in terms of the temperature relative to the Fermi energy ($T/\varepsilon_f$) which in turn was parametrized in terms of the normalized multiplicity fluctuation. The nucleon density $\rho$ is therefore determined from the Fermi-energy relation $\varepsilon_f=\varepsilon_{f_0}(\rho/\rho_0)^{2/3}$ with $\varepsilon_{f_0}$ and $\rho_0$ respectively the ground-state values of Fermi energy and nucleon density. Coulomb corrections were later applied to derived temperatures and densities. This is done by a method borrowed from electron scattering where the Coulomb field is taken to be the Fourier transform of the Coulomb potential of the source. In such an approach, as described in Refs.~\cite{PhysRevC.88.024607,JPGNuclPartPhys.41.055109}, the equations of quadrupole momentum fluctuation, the average multiplicity, as well as the multiplicity fluctuation which contain the Coulomb field term are numerically solved to derive the temperature $T$, the density $\rho$, and the volume $V$ of the system. This method can be considered reliable as from model calculations, temperatures of protons after Coulomb corrections were similar to those of neutrons. In Ref.~\cite {PhysRevC.90.027602}, where we have applied this method to experimental data, it was observed that these Coulomb corrections lower temperature values by almost 2 MeV while their effects were small on derived densities. The error in applying the Coulomb corrections arises from the uncertainty in the source charge. We varied the source charge by $\pm$ 2 units and the estimated errors are respectively $\pm2\%$ for the densities and $\pm6\%$ for the temperatures. 

Previously \cite{J.Phys.Conf.Ser.420.012110,PhysRevC.87.017603,IntJModPhysE.22.1350090} the data were sorted into four different QP asymmetry bins of width 0.05, ranging from 0.04 to 0.24. In a subsequent paper \cite{PhysRevC.90.027602}, it was shown that values of $T$ and $\rho$ that correspond to an asymmetry bin width close to zero (as it should be for fixed $A$ and $Z$) could be obtained by averaging values for all four asymmetry bins. Therefore, in this paper we have reported averaged values of temperatures and densities for all four asymmetry bins. On the same grounds, $F_{sym}$ values presented here are also obtained from averaged values of Landau's equation fitting parameters from the four asymmetry bins.

\section{RESULTS AND DISCUSSION}

\begin{figure}[!htp]
\includegraphics[width=8.5cm,height=7.5cm,angle=0]{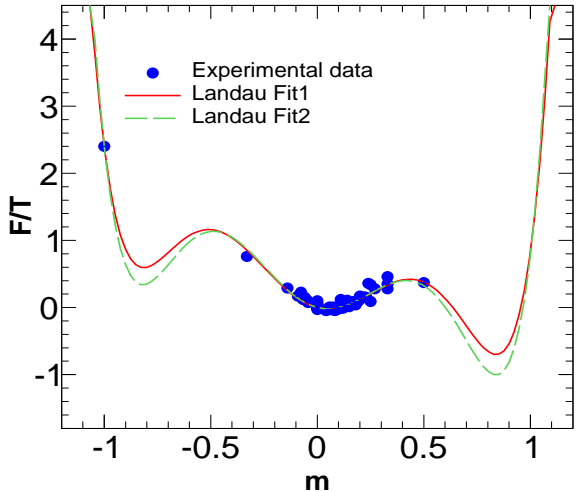}
\caption{(Color online) $F/T$ values as a function of fragment's
neutron-proton asymmetry $m$ for an excitation energy of 5.5 MeV/A of the QP. The solid line (Landau Fit1) is a fit to the data with all Landau's equation parameters as free parameters while the dashed line (Landau Fit2) represents a fit to data fixing $c=115$. Error bars corresponding to statistical errors are smaller than the symbols.}
\label{Fig1}
\end{figure}

Figure~\ref{Fig1} shows the free energy ($F/T$) values as a function of fragment's neutron-proton asymmetry $m$ at an excitation energy of 5.5 MeV/A of the QP. As the efficiency for measuring neutrons differs from the efficiency for measuring charged particles, only charged-particle yields are used in the Landau's equation fitting. The form of the fitting function has physical restrictions on it. Since the symmetry energy is isospin symmetric, only even powers of $m$ appear, aside from the external field. Furthermore, at extreme values of asymmetry, the free energy must not be decreasing toward negative infinity. Within these constraints, the form must also have a minimum at zero as dictated by the data. A parabolic fit describes the data only around zero, but misses entirely the points at both $m=-1$ and $m=0.5$. Our ability to measure this point is very useful to constrain $F/T$. Having excluded the quadratic fit, we next rule out all quartic fits immediately on the grounds that the function must be rising at extreme values of $m$ and still have a minimum at 0. A sixth order (in even terms only) is the next simplest polynomial that satisfies the physical constraints. The values of the free energy obtained were corrected for pairing \cite {PhysRevLett.101.122702,PhysRevC.87.017603}, similar to the mass formula, and a good scaling is seen in the figure. The solid line (Landau Fit1) is a fit to the data with all parameters as free parameters. In Ref.~\cite {PhysRevC.87.017603}, the parameter $c$ was observed to be almost constant, within uncertainties, over the entire range of the QP excitation energy. Here we additionally fitted the data fixing $c=115$. This is represented by the dashed line (Landau Fit2). It is observed that the two fitting curves provide a good fit to the free-energy data. However, at the two extreme minima (at large $m$ values) where we have no data points the two curves are slightly different. The values of $a$, $b$ and $c$ corresponding to the solid line were obtained as 15.527$\pm$0.041, -91.786$\pm$0.484 and 100.81$\pm$0.615, respectively. After fixing $c=115$ (dashed line), the values of $a$ and $b$ were obtained as 16.289$\pm$0.024 and -102.871$\pm$0.058, respectively. This resulted in similar values of $F_{sym}/T$, calculated using Eq.~\ref{Eq4}, as 1.619$\pm$0.16 and 1.593$\pm$0.019. The values of parameters $a$ and $b$ of Landau's equation used in the rest of the discussion were obtained by fixing $c=115$. In this way, estimated errors on extracted $F_{sym}$ values were significantly minimized. The appearance of the three minima is a signature of a first-order phase transition of the system \cite{PhysRevLett.101.122702,PhysRevC.81.044618,PhysRevC.83.054609,S021830131250019X,J.Phys.Conf.Ser.420.012110,PhysRevC.87.017603}.

As there are no experimental data points in the region of the two minima of the fit and one must rely on the proton point at $m=-1$, we have investigated the uncertainty on the $F_{sym}$ values introduced by the fit parameters if this point was in error by some amount. A $15\%$ variation of the $F/T$ value at $m=-1$ resulted in a $19\%$ change of the $F_{sym}$ value. We estimated the systematic uncertainty caused by the isolated proton point on the $F_{sym}$ values to be less than 20$\%$.

\begin{figure}[!htp]
\includegraphics[width=9.5cm,height=7.5cm,angle=0]{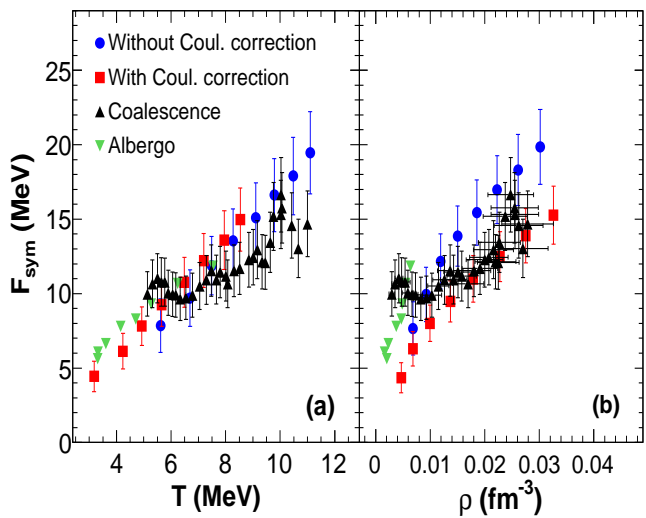}
\caption{(Color online) (a) Symmetry free energy vs temperature. (b) Symmetry free energy vs density. The symmetry free energy coefficients ($F_{sym}$) are extracted from Landau's free energy approach. Temperatures and densities are derived from the quantum-fluctuation method with protons as the probe particle. Full circles and squares correspond respectively to results without and with Coulomb correction. For comparison, $F_{sym}$ values obtained from isoscaling techniques along with $T$ and $\rho$ from Albergo \cite{PhysRevLett.104.202501} and coalescence \cite{PhysRevC.85.064618} methods are also plotted. Statistical errors are indicated by the bars and are not shown when smaller than the symbols.}
\label{Fig2}
\end{figure}

The symmetry free energy coefficients determined from the Landau free-energy technique are displayed as a function of the system temperature $T$ (Fig.~\ref{Fig2}(a)) and density $\rho$ (Fig.~\ref{Fig2}(b)). $T$ and $\rho$ for this work were derived from the quantum-fluctuation method without and with Coulomb corrections, while Fsym values are derived from the Landau equation parameters. The present data set is compared to previously published work, which used the Albergo method. The Albergo method is a double ratio technique that evaluates the temperature and density of equilibrated nuclear regions using the yields of different light nuclides ($d$, $t$, $h$, $\alpha$). Application of this technique assumes that thermal equilibration and chemical equilibration have been attained \cite{Albergo,PhysRevC.75.014601}. Data represented by inverted triangle symbols (Albergo) are taken directly from Ref.~\cite{PhysRevLett.104.202501}, where $T$ and $\rho$ were determined. The isoscaling method was used to extract $F_{sym}$ values. A comparison is also made to previous work that extracted $T$ and $\rho$ using a combination of the Albergo method and a coalescence model. This work is shown in full up-triangle symbols, the points are taken directly from Ref.~\cite{PhysRevC.85.064618}. The overlap of the errors from this work reflect that the binning of the data was not done as a function of temperature, density, or free symmetry energy, but of fragment surface velocity, which is correlated to the emission time. The effect of this is seen in the size and proximity of the error bars. In the coalescence model, the momentum space densities of ejected light composite particles are directly related to those of the ejected nucleons of the same velocity. The phase space correlations, which lead to cluster formation, may therefore be parametrized in terms of the radius of the momentum space volume, $P_0$, within which the correlations exist. In this way the double isotope yield ratio at equal velocity is used to determine the temperature. From the relationship between the coalescence parameter $P_0$ and the volume of the emitting system, the density is derived \cite{PLB473.29,PhysRevC.62.034607}. Isoscaling was also used to extract $F_{sym}$ in this case.

It is observed that there is a fair agreement between temperature-dependent $F_{sym}$ results. However, the density-dependent $F_{sym}$ results with Coulomb corrections significantly deviate from those obtained without Coulomb corrections and Coalescence and Albergo data as well. Neverthless, it is amazing to see that within the error bars temperature- and density-dependent symmetry free energies derived from different methods agree with each other to a remarkable degree. In Ref.~\cite{PhysRevC.85.064618}, where the colescence data have been taken, the quoted errors on the temperatures are $10\%$ at low density evolving to $15\%$ at the higher densities. The error in the derivation of the density was estimated to be in the order of $17\%$.  

\begin{figure}[!h]
\includegraphics[width=9.5cm,height=7.cm,angle=0]{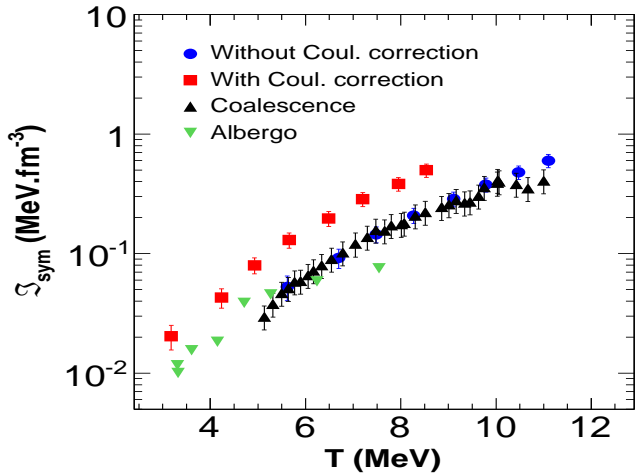}
\caption{(Color online) Symmetry free-energy density ($\mathfrak{J}_{sym}=F_{sym}\times\rho$) as a function of the system temperature $T$. Error bars represent statistical errors and are not shown when smaller than the symbols.}
\label{Fig3}
\end{figure}

From the values of $F_{sym}$, $T$ and $\rho$, we examine in Fig.~\ref{Fig3} the symmetry free-energy density ($\mathfrak{J}_{sym}=F_{sym}\times\rho$) against $T$. It is observed that $\mathfrak{J}_{sym}$ monotonically increases as $T$ increases and some of the differences seen between curves in both panels of Fig.~\ref{Fig2} are less evident except for the curve represented by full squares where Coulomb corrections have been accounted for. Therefore, plots of $F_{sym}$ values as a function of $T(\rho)$ could be misleading, since $T$ and $\rho$ vary simultaneously. The $\mathfrak{J}_{sym}$ as a function of temperature makes use of all the three quantities that are experimentally accessed. This quantity displays a clear deviation among the different methods especially when dealing with the Coulomb correction, suggesting some model dependence in dealing with the Coulomb term. This ambiguity could be further studied by deriving density and temperature from neutron fluctuations as well. We expect from model calculations \cite{JPGNuclPartPhys.41.055109} that quantities derived using the protons should be similar to those of neutrons after the Coulomb correction. If this is not true, then further work is needed to understand the role of Coulomb. In the coalescence model a Coulomb correction is applied through a shift of the measured kinetic energy spectra of the particles of interest. From such a Coulomb shift, knowing the charge and mass of the emitting system, it is possible to derive its density \cite{PhysRevLett.88.042701} which can be compared to the obtained values from coalescence \cite{PhysRevC.85.064618,PhysRevLett.108.172701}. If the densities obtained do not agree with each other, then a simple shift of the energy distribution is not sufficient to describe the role of Coulomb.

\section{CONCLUSIONS}
In conclusion, we have shown that symmetry free energy coefficients of nuclear systems can be extracted from fragment yield data produced in Fermi-energy heavy-ion collisions by employing Landau's free-energy approach. The temperature- and density-dependent symmetry free energies have been observed to be consistent with those derived from isoscaling analyses. This is the first time experimental fragment yield data analyzed within the Landau description have been used to determine symmetry free energy coefficients. We have found some discrepancy among different methods possibly because of the different handling of Coulomb corrections. Precise measurement of the neutron distribution function might help to solve this ambiguity. The estimation of entropic contributions to the symmetry free energy in order to derive symmetry energy coefficients is currently being given special attention. 

\begin{acknowledgements}
This work was supported by the Robert A. Welch Foundation under Grant No. A-1266 and the U. S. Department of Energy under Grant No. DE-FG03-93ER-40773.
\end{acknowledgements}

\bibliography{Landau_Fsym}

\end{document}